\documentclass[pra,twocolumn,superscriptaddress,preprintnumbers,amsmath,amssymb]{revtex4}
\usepackage[latin1]{inputenc}
\usepackage{bm}
\usepackage{multirow,amssymb,amsbsy,amsmath,amsfonts}
\usepackage{graphicx}
\usepackage{verbatim}
\makeatletter
\usepackage{pifont}
\newcommand{\tr}[0]{\text{tr}}
\def\bra#1{\mathinner{\langle{#1}|}}
\def\ket#1{\mathinner{|{#1}\rangle}}

\usepackage{color}

\makeatother

\begin{document}
\title{Experimental realization of high-fidelity teleportation via non-Markovian open quantum system}

\author{Zhao-Di Liu}
\affiliation{CAS Key Laboratory of Quantum Information, University of Science and Technology of China, Hefei 230026, China}
\affiliation{CAS Center For Excellence in Quantum Information and Quantum Physics, University of Science and Technology of China, Hefei 230026, China}

\author{Yong-Nan Sun}
\affiliation{CAS Key Laboratory of Quantum Information, University of Science and Technology of China, Hefei 230026, China}
\affiliation{CAS Center For Excellence in Quantum Information and Quantum Physics, University of Science and Technology of China, Hefei 230026, China}

\author{Bi-Heng Liu}
\affiliation{CAS Key Laboratory of Quantum Information, University of Science and Technology of China, Hefei 230026, China}
\affiliation{CAS Center For Excellence in Quantum Information and Quantum Physics, University of Science and Technology of China, Hefei 230026, China}

\author{Chuan-Feng Li}
\email{cfli@ustc.edu.cn}
\affiliation{CAS Key Laboratory of Quantum Information, University of Science and Technology of China, Hefei 230026, China}
\affiliation{CAS Center For Excellence in Quantum Information and Quantum Physics, University of Science and Technology of China, Hefei 230026, China}

\author{Guang-Can Guo}
\affiliation{CAS Key Laboratory of Quantum Information, University of Science and Technology of China, Hefei 230026, China}
\affiliation{CAS Center For Excellence in Quantum Information and Quantum Physics, University of Science and Technology of China, Hefei 230026, China}

\author{Sina Hamedani Raja}
\affiliation{Turku Centre for Quantum Physics, Department of Physics and 
Astronomy, University of
Turku, FI-20014 Turun yliopisto, Finland}

\author{Henri Lyyra}
\affiliation{Turku Centre for Quantum Physics, Department of Physics and 
Astronomy, University of
Turku, FI-20014 Turun yliopisto, Finland}
\affiliation{Department of Physics and Nanoscience Center, 
University of Jyv\"askyl\"a, FI-40014 University of Jyv\"askyl\"a, Finland}

\author{Jyrki Piilo}
\email{jyrki.piilo@utu.fi}
\affiliation{Turku Centre for Quantum Physics, Department of Physics and 
Astronomy, University of
Turku, FI-20014 Turun yliopisto, Finland}

\date{\today}

\begin{abstract}
Open quantum systems and study of decoherence are important for our fundamental understanding of quantum physical phenomena. 
For practical purposes, there exists a large number of quantum protocols exploiting quantum resources, e.g. entanglement, which allows to go beyond what is possible to achieve by classical means. We combine concepts from open quantum systems and quantum information science, and give a proof-of-principle experimental demonstration -- with teleportation -- that it is possible to implement efficiently a quantum protocol via non-Markovian open  system. The results show that, at the time of implementation of the protocol, it is not necessary to have the quantum resource in the degree of freedom used for the basic protocol -- 
%as long as there exists useful resources including some other degree of freedom. 
as long as there exists some other degree of freedom, or environment of an open system, which contains useful resources.
The experiment is based on a pair of photons, where their polarizations act as open system qubits and frequencies as their environments --  while the path degree of freedom of one of the photons represents the state of Alice's qubit to be teleported to Bob's polarization qubit.

 \end{abstract}

\maketitle

%\textit{Introduction.---}% 
\section{Introduction} 

The study of open quantum systems is important both for fundamental and practical purposes. When an open system interacts with its environment, this typically leads to decoherence and loss of quantum properties~\cite{Breuer2007,Rivas2012}, which -- in turn -- makes it often difficult to implement quantum protocols in an ideal manner in experiments~\cite{Suter2016}.
In general, during the last ten years, there have been significant developments in understanding and characterizing the abundant and diverse features of
 open system dynamics~\cite{RevRivas,Breuer2016,devega,LiRev,p1,p2}.
These developments have influenced, and have been influenced by, the increasing ability to realize experimentally reservoir engineering~\cite{wineland}, various fundamental open system models in non-Markovian regime~\cite{exp2},  and the control of open system dynamics~\cite{zdliu}. Indeed, by now a number of various physical platforms have been used for this purpose
including, e.g.,  optical systems~\cite{exp2,zdliu,achiu,bhliu13,fff,Ber2015,sdc,scia1,syu,acu,sau}, NV-centers~\cite{haa,nori2020}, trapped ions~\cite{mwit}, and NMR-systems~\cite{Ber2016,dku}. In addition to fundamental studies and tests, recent experimental work also includes some of the first exploitations of non-Markovian memory effects in basic quantum information protocols including, e.g., single qubit Deutsch-Josza algorithm~\cite{DJ}.

Considering entanglement, as a matter of fact, it plays a dual role when considering implementation of  quantum protocols with systems interacting with their environments. To start with, we need entanglement -- as a quantum resource -- when implementing quantum protocols and to go beyond what can be achieved by classical resources. However, when an open system interacts with its environment, the entanglement within the open system decreases due to decoherence,  and the efficiency of the quantum protocol typically diminishes. At the same time, the open system often gets entangled with its environment. This means that the total system-environment state still contains useful  resource while it does not anymore reside in the degrees of freedom which are explicitly used for the implementation of the quantum protocol.

Therefore, we arrive to the following question: Is it possible to make efficient experimental realization of a quantum information protocol via open quantum system?
We answer this question affirmatively and demonstrate a proof-of-principle experiment by using teleportation~\cite{Bennett} as an example. 
This also means that we combine the central concepts of quantum information and open quantum systems in a new fundamental manner in an experiment.
Note that for technological purposes, there has been recently impressive experiments, e.g., demonstrating ground-to-satellite teleportation~\cite{g2sat}
and superdense teleportation for information transfer using a pair of hyper-entangled photons~\cite{Kwiat2015}. 
%Even though our current results in principle could be used to search for avenues for improving practical implementations in noisy conditions, too, 
We, instead, rather see our current contribution dealing with fundamental questions on open quantum systems and quantum information. Our experiment uses the concept of nonlocal memory effects~\cite{nlnm,bhliu13,sdc} and a recent theoretical proposal how to exploit them in teleportation~\cite{telep}. For completeness, we next recall the basic steps of the scheme, including some changes compared to the original theoretical proposal, and then continue to the experimental part and results.

%\textit{Theoretical description.---}%
\section{Theoretical description}

First, we prepare a polarization entangled state $\ket{\phi^+}_{ab}=\frac{1}{\sqrt{2}}( \ket{HH}+\ket{VV})$, where $H(V)$ denotes the horizontal (vertical) polarization of the photon.
The total initial polarization-frequency two-photon state is
\begin{equation}\label{eq:1}
\ket{\psi(0)}=\ket{\phi^+}_{ab}\otimes \int{d\omega_a d\omega_b g(\omega_a,\omega_b)\ket{\omega_a}\ket{\omega_b}},
\end{equation}
where $g(\omega_a,\omega_b)$ is the joint frequency amplitude distribution of the photons $a$ and $b$ with $\int d\omega_a d\omega_b|g(\omega_a,\omega_b)|^2=1$. 
We describe further down below in more detail what role the properties of $|g(\omega_a,\omega_b)|^2$ play in the protocol.
Before Alice receives her photon, its polarization and frequency are coupled in a quartz plate. This local interaction is given by the time-evolution operator
\begin{equation}
U(t) |\lambda \rangle |\omega\rangle =  \exp(i n_{\lambda} \omega t) |\lambda\rangle |\omega\rangle,
 \end{equation}
 where $\lambda$ denotes the given polarization direction 
and $n_{\lambda}$ its index of refraction in the quartz plate. Even though the polarization-frequency state remains pure, this leads to dephasing of the polarization degree of freedom~\cite{exp2,bhliu13}. Indeed, after the local interaction in the side of Alice, the two-photon polarization-frequency state is both pure and entangled
\begin{equation}
\ket{\psi(t_a)}=\frac{1}{\sqrt{2}}(\ket{HH}\ket{\xi_{HH}(t_a)}+\ket{VV}\ket{\xi_{VV}(t_a)}), 
\end{equation}
where $\ket{\xi_{\lambda \lambda}(t_a)}=\int d\omega_a d \omega_b g(\omega_a,\omega_b)e^{i n_{\lambda}^a\omega_a t_a}\ket{\omega_a}\ket{\omega_b}$.
The joint polarization state, in turn, is not anymore fully entangled and has become a mixed state 
\begin{eqnarray}\label{eq:3}
\rho_{ab}(t_a)&=&\frac{1}{2}(\ket{HH}\bra{HH}+\kappa_1(t_a)\ket{HH}\bra{VV} \nonumber\\
&&+\kappa_1^*(t_a)\ket{VV}\bra{HH}+\ket{VV}\bra{VV}),
\end{eqnarray}
where the local decoherence function is
\begin{equation}
\label{eq6}
\kappa_a(t_a)=\int d\omega_ad\omega_b |g(\omega_a,\omega_b)|^2 e^{-i \Delta n_a \omega_a t_a}
\end{equation}
with $\Delta n_a=n_V^a-n_H^a$. 

Alice prepares now a third qubit in a state, that she wants to teleport to Bob. Unlike the original proposal where a third photon is used for this purpose~\cite{telep},
she introduces binary path degree of freedom of the photon she possesses.
In general this path-qubit state is 
%\begin{equation}
$\ket{\phi}_s=\alpha\ket{0}+\beta\ket{1}$,
%\end{equation}
where $|0\rangle$ and $|1\rangle$ denote the two spatial paths she uses. 
Therefore, Alice's and Bob's over-all state is now 
$\ket{\Psi(t_a)}=\frac{1}{\sqrt{2}}\ket{\phi}_s\left[(\ket{HH}\ket{\xi_{HH}(t_a)}+\ket{VV}\ket{\xi_{VV}(t_a)}\right].$ 
We can now write this state by using the Bell-state basis of the two qubits of Alice and obtain
\begin{eqnarray} \label{eq:6}
\ket{\Psi(t_a)}&=&\frac{1}{2}\ket{\Phi^+}_{sa} [\alpha\ket{H}_b\ket{\xi_{HH}(t_a)}+\beta\ket{V}_b\ket{\xi_{VV}(t_a)}]\nonumber\\
&+&\frac{1}{2}\ket{\Phi^-}_{sa}  [\alpha\ket{H}_b\ket{\xi_{HH}(t_a)}-\beta\ket{V}_b\ket{\xi_{VV}(t_a)}]\nonumber\\
&+&\frac{1}{2}\ket{\Psi^+}_{sa} [\beta\ket{H}_b\ket{\xi_{HH}(t_a)}+\alpha\ket{V}_b\ket{\xi_{VV}(t_a)}]\nonumber\\
&+&\frac{1}{2}\ket{\Psi^-}_{sa} [\alpha\ket{V}_b\ket{\xi_{VV}(t_a)}-\beta\ket{H}_b\ket{\xi_{HH}(t_a)}],\nonumber \\
\end{eqnarray}
where $\ket{\Phi^{\pm}}_{sa}=\frac{1}{\sqrt{2}}( \ket{0}\ket{H}\pm\ket{1}\ket{V})$ and $\ket{\Psi^{\pm}}_{sa}=\frac{1}{\sqrt{2}}( \ket{0}\ket{V}\pm\ket{1}\ket{H})$. 
It is worth noting that in each line -- corresponding to four outcomes of Alice's measurement -- we have a pure and entangled state between Bob's polarization and the frequencies of the two-photons, in addition to the amplitudes $\alpha$ and $\beta$ being transferred.

Alice communicates her measurement result to Bob. What should he do now? Bob first applies one of the four unitary transformations on his qubit -- according to the standard teleportation scheme -- and after this applies local polarization-frequency dephasing interaction to his photon.
The cases corresponding to four outcomes of Alice are
(i) $\ket{\Phi^+}_{sa} \Rightarrow \mathbb{I}, \quad \Delta n_b=\Delta n_a$;
(ii) $\ket{\Phi^-}_{sa} \Rightarrow  \sigma_z, \quad \Delta n_b=\Delta n_a$;
(iii) $\ket{\Psi^+}_{sa} \Rightarrow \sigma_x, \quad \Delta n_b=-\Delta n_a$;
(iv) $\ket{\Psi^-}_{sa} \Rightarrow i \sigma_y, \quad \Delta n_b=-\Delta n_a$.
%\begin{equation} \label{eq:7}
%\begin{array}{ccc}
%\ket{\Phi^+}_{sa}& \Rightarrow& \mathbb{I}, \quad \Delta n_b=\Delta n_a \\
%\ket{\Phi^-}_{sa} &\Rightarrow & \sigma_z, \quad \Delta n_b=\Delta n_a \\
%\ket{\Psi^+}_{sa}& \Rightarrow& \sigma_x, \quad \Delta n_b=-\Delta n_a \\
%\ket{\Psi^-}_{sa} &\Rightarrow& i \sigma_y, \quad \Delta n_b=-\Delta n_a.
%\end{array}
%\end{equation}
Here, $\sigma_x, \sigma_y$, and $\sigma_z$ are the unitary Pauli rotations, and $\Delta n_b$ indicates the conditional choice for Bob's  birefringence. Let us as an example to check one of the four cases in detail. 

Suppose that Alice's measurement outcome was  $\ket{\Phi^+}_{sa}$ corresponding to the first line of Eq.~(\ref{eq:6}).
For this case, Bob's unitary qubit transformation is the indentity operator and he only need to apply dephasing noise with $\Delta n_b=\Delta n_a$.
Once Bob applies the dephasing interaction for the duration $t_b$, the state of his polarization qubit is
\begin{eqnarray} \label{eq:10}
\rho_b &=&|\alpha|^2 \ket{H}\bra{H}+\alpha\beta^*\kappa(t_a,t_b) \ket{H}\bra{V}\\
&&+\alpha^*\beta \kappa^*(t_a,t_b) \ket{V}\bra{H}+|\beta|^2 \ket{V}\bra{V}, \nonumber
\end{eqnarray}
where the decoherence function $\kappa(t_a,t_b)$ is given by 
\begin{equation}
\kappa(t_a,t_b)=\int d \omega_a d \omega_b |g(\omega_a,\omega_b)|^2 e^{-i \Delta n_b (\omega_a t_a+\omega_b t_b)}.
\end{equation}
Note that the photon on the side of Alice is already destroyed. However, the influence of Bob's local noise on his polarization state does depend on the initial joint two-photon frequency distribution $|g(\omega_a,\omega_b)|^2$. 

The question now becomes whether it is possible to have $|\kappa (t_a,t_b)|=1$ so that Bob eventually has pure polarization state after his noise -- whilst in all of the previous points of the protocol the state has been mixed. For this purpose, let us study the properties of the initial two-photon frequency distribution
 $|g(\omega_a,\omega_b)|^2$.
 
We consider the joint two-photon frequency distribution  $|g(\omega_a,\omega_b)|^2$ in a downconversion process as a bivariate Gaussian distribution~\cite{nlnm,bhliu13} with covariance matrix elements  $C_{ij}=\langle\omega_i\omega_j
\rangle-\langle\omega_i\rangle\langle\omega_j\rangle$.
The means and variances for the two photons are 
$\langle \omega_a \rangle = \langle \omega_b \rangle =\omega_0/2$, where $\omega_0$ is the frequency of the downconversion pump, and $C_{11} = 
C_{22} = \langle\omega^2_i\rangle-\langle\omega_i\rangle^2$. The frequency-frequency correlation is quantified by the coefficient $K=C_{11}/\sqrt{C_{11}
C_{22}}=C_{12}/C_{11}$, such that $|K|\leq1$. 
Taking initially a maximally anti-correlated frequency distribution with $K=-1$ having $\omega_a+\omega_b=\omega_0$,
and Bob using interaction time $t_b=t_a$, the magnitude of  his decoherence function becomes $|\kappa (t_a,t_b)|=1$ and he obtains -- after applying the local noise --  the pure polarization state   
\begin{equation} \label{eq:12}
|\psi_F\rangle=\alpha |H\rangle+\beta e^{i \omega_0 \Delta n_b t_b} |V\rangle.
\end{equation}
Bob can in straightforward manner cancel the extra relative phase with $\omega_0$, and therefore succeeds in the teleportation with fidelity equal to one.
Note that in general for dephasing, the width of the single-peak Gaussian frequency distribution defines how strongly the exponential damping of the  magnitude of the decoherence function occurs -- when having a single photon or noise only on one side of the two-photon system destroying the polarization entanglement  (see, e.g., Ref.~\cite{exp2}). In the ideal case described above in Eq.~(\ref{eq:12}), both photons still have finite local frequency distribution widths ($C_{11}$ and $C_{22}$) and subsequent source of dephasing even though the final teleported state can be pure state due to having $K=-1$.  
%All the other remaining  cases corresponding to three other measurement outcomes of Alice go in the same manner.

%\textit{Expererimental set-up.---}%
\section{Experimental set-up}

\begin{figure}[t!]
\begin{center}
  \includegraphics[keepaspectratio,width=0.45\textwidth]{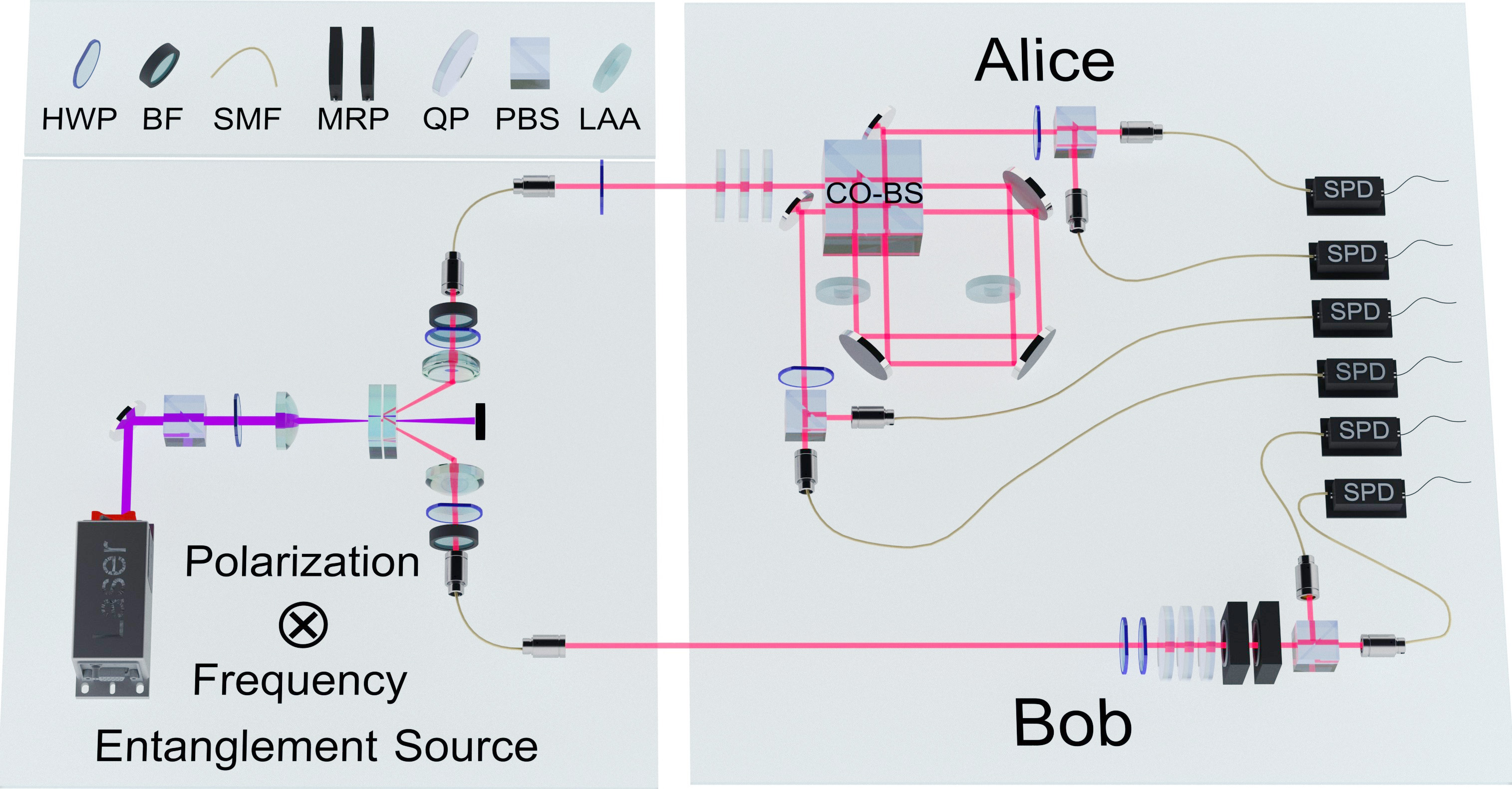}
  \caption{\label{fig:set-up}  Experimental set-up. The two-photon state is prepared by the spontaneous parametric down conversion (SPDC) pumping the beta-barium-borate (BBO) crystal with a continuous wave (CW) laser. The BFs  are used to choose the photons.
QPs implement the noise in Alice's side. To prepare Alice's path qubit state and to prepare for path-qubit Bell measurement, a specifically designed Sagnac interference ring
with very high stability is used.
Here, we use a combined beam splitter, abbreviated with CO-BS, which consists of beam splitter and polarizing beam splitter parts.
 The path qubit state itself is prepared with LAAs along the paths inside the interferometer.
 Alice's HWPs, PBSs and SPDs complete the Bell-measurement. In Bob's side, HWPs induce unitary transformation, QPs dephasing noise, and state tomography 
is performed with MRP, PBS and SPDs.
Abbreviations: HWP: half wave plate, BF: bandpass filter, SMF: single mode fiber, MRP: motor rotating plate, QP: quartz plate, PBS: polarizing beam splitter, LAA: linear adjustable attenuator, SPD: single photon detector.
}
  \end{center}
\end{figure}

%The experiment consists of the following steps: (i) preparation of a two-photon state which is entangled in polarization and in frequency while the polarization and frequency are not correlated between them; (ii) dephasing noise for the polarization on Alice's side; (iii) preparation of Alice's path qubit state to be teleported; (iv) Alice's path-polarization Bell-measurement; (iv) Bob having Alice's measurement outcome applies unitary operation on his polarization qubit; (v) Bob applies dephasing noise to his polarization qubit; (vi)  state tomography of Bob's qubit.
We display the experimental set-up in Fig.~\ref{fig:set-up}.
To prepare the required initial state, we exploit SPDC process  where a $404$nm CW laser pumps a BBO crystal. This crystal is made of two orthogonal glued type-I phase matched BBOs and can be used to produce polarization entangled photons by using the $H+V$ direction polarized laser pump.
The frequency correlations, in turn, arise due to the narrow linewidth of the CW pump laser.
In our case, the $404$nm pump has measured linewidth below $0.06$nm.
After downconversion, the photons have a bandwidth on the order of $135$nm. Then, $3$nm full width at half maximum (FWHM) bandpass filters (centered at $808$ nm) are used to choose the most indistinguishable SPDC photon pairs. Even though the filtering changes the frequency distribution, there is still high amount of frequency-frequency correlations left due the very narrow pump linewidth.
Note also that in the used $3$nm frequency window, the difference between the indices of refraction for ordinary and extraordinary rays can be considered constant with value $0.00889$.
%
 %The frequency correlations, in turn, arise due to the narrow linewidth of the CW pump laser.
%Then, $3$nm full width at half maximum (FWHM) bandpass filters (centered at $808$ nm) are used to choose the most indistinguishable SPDC photon pairs. 

On Alice's side, we first insert quartz plates, with increasing thickness corresponding to increasing interaction times,  to induce dephasing noise for Alice's polarization qubit. 
Then Alice prepares her path qubit to be teleported, and makes the polarization-path Bell-measurement. Earlier proposals for path-polarization measurement have used, e.g., Mach-Zehnder  interferometer~\cite{zender}. However, to improve significantly the stability of the scheme required for teleportation, we have designed a specific Sagnac interferometer for this purpose.
Here, the crucial component is a specific beam splitter (CO-BS) which consists half of BS and half of PBS.
When the photon enters the Sagnac interferometer, it goes through the BS part of the CO-BS. Along the paths of the interferometer, there are LAAs
 that can produce arbitrary ratio of the two paths and prepares the path qubit state to be teleported. Note that the path degree of freedom is fully independent of the polarization and frequency degrees of freedom of the photon. 
Alice can then make the Bell measurement by interfering her two paths, when the photon exits the interferometer in the PBS part of CO-BS, and by using other PBS at $45^{o}$ at each output path before the photon hits the SPDs at the outputs.

On Bob's side, he first implements with HWP the unitary operation on his polarization qubit based on Alice's Bell-measurement outcome. After this, he induces dephasing noise by using quartz plates to couple the polarization (open system) and frequency (environment). Finally, using MRPs and PBS, he performs the state tomography of his qubit and completes the protocol.

%\textit{Results.---}%
\section{Results}

We present now two sets of experimental results.
In both cases, the success of the teleportation is quantified in usual manner by  fidelity 
$F(\rho_{\text{out}}, \rho_{\text{in}}) = \left(\tr \sqrt{\sqrt{\rho_{\text{out}}} \rho_{\text{in}}
\sqrt{\rho_{\text{out}}}} \right)^2$ between the state Alice prepared $\rho_{\text{in}}$ and  $\rho_{\text{out}}$ that Bob measured after completing the protocol.

%The theoretical expression of the fidelity in the scheme is $F=1-2|\alpha|^2|\beta|^2 (1-|\kappa_a(t_a)|^{2\delta K})$ with $K=-1+\delta K$~\cite{telep}.
Alice's qubit resides in path degree of freedom of the photon while Bob's qubit corresponds to polarization state of another photon. 
Note that the maximum fidelity with fully classical mixed states and using LOCC is $2/3$~\cite{verst}.
The results are presented for all possible four Bell-state measurement outcomes of Alice $\{|\Phi^+\rangle,~|\Phi^-\rangle,~|\Psi^+\rangle,~|\Psi^-\rangle\}$.

 \begin{figure}[t!]
\begin{center}
  \includegraphics[keepaspectratio,width=0.45\textwidth]{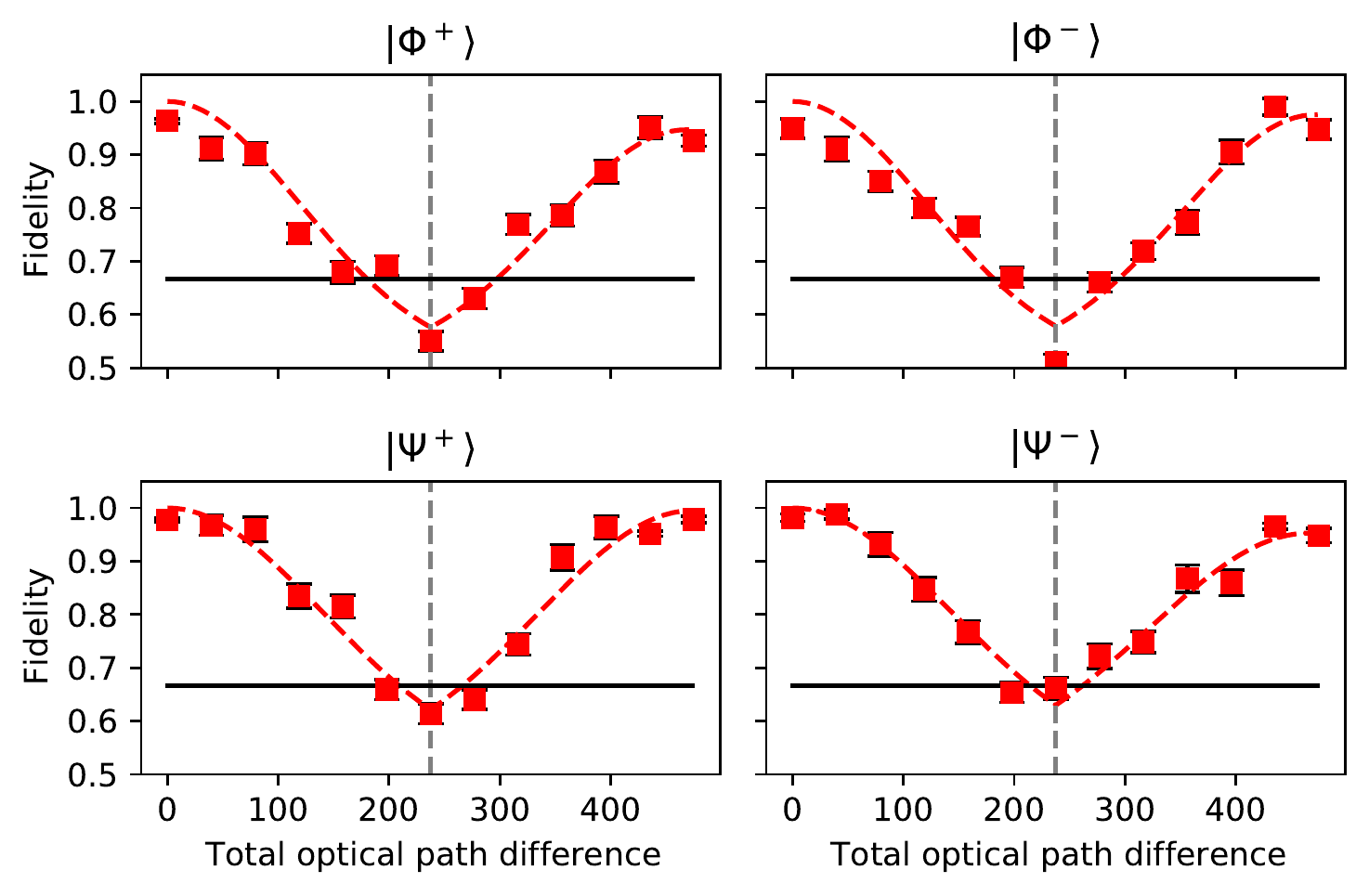}
  \caption{\label{fig:res1} The measured teleportation fidelity with error bars as a function of amount of noise first on Alice's side and followed by increasing amount of noise on Bob's side. The vertical lines indicate the sides of Alice and Bob for the noise. The dashed line is the theoretical fit for the experimental results including the possibly non-ideal value of $K$~\cite{telep}.
 The fits give an estimate $-0.997 \leqslant K \leqslant -0.963$.
  Teleported input state is $\rho_{\text{in}}=\rho_{+}$ [Eq.~\eqref{eq:in1}]. The results correspond to the first set of experimental results (see the main text). The optical path difference is expressed with the unit of $808$nm. When Alice increases noise on her side, the fidelity decreases. With increasing amount of noise on Bob's side, the fidelity recovers. The black horizontal line indicates the classical limit of the fidelity with value of $2/3$.
The error bars are standard deviations calculated by the Monte-Carlo method and mainly due to the counting statistics.
In  most cases, the error bars are smaller than the symbols. }
  \end{center}
\end{figure}
 
 \begin{figure}[t!]
\begin{center}
  \includegraphics[keepaspectratio,width=0.45\textwidth]{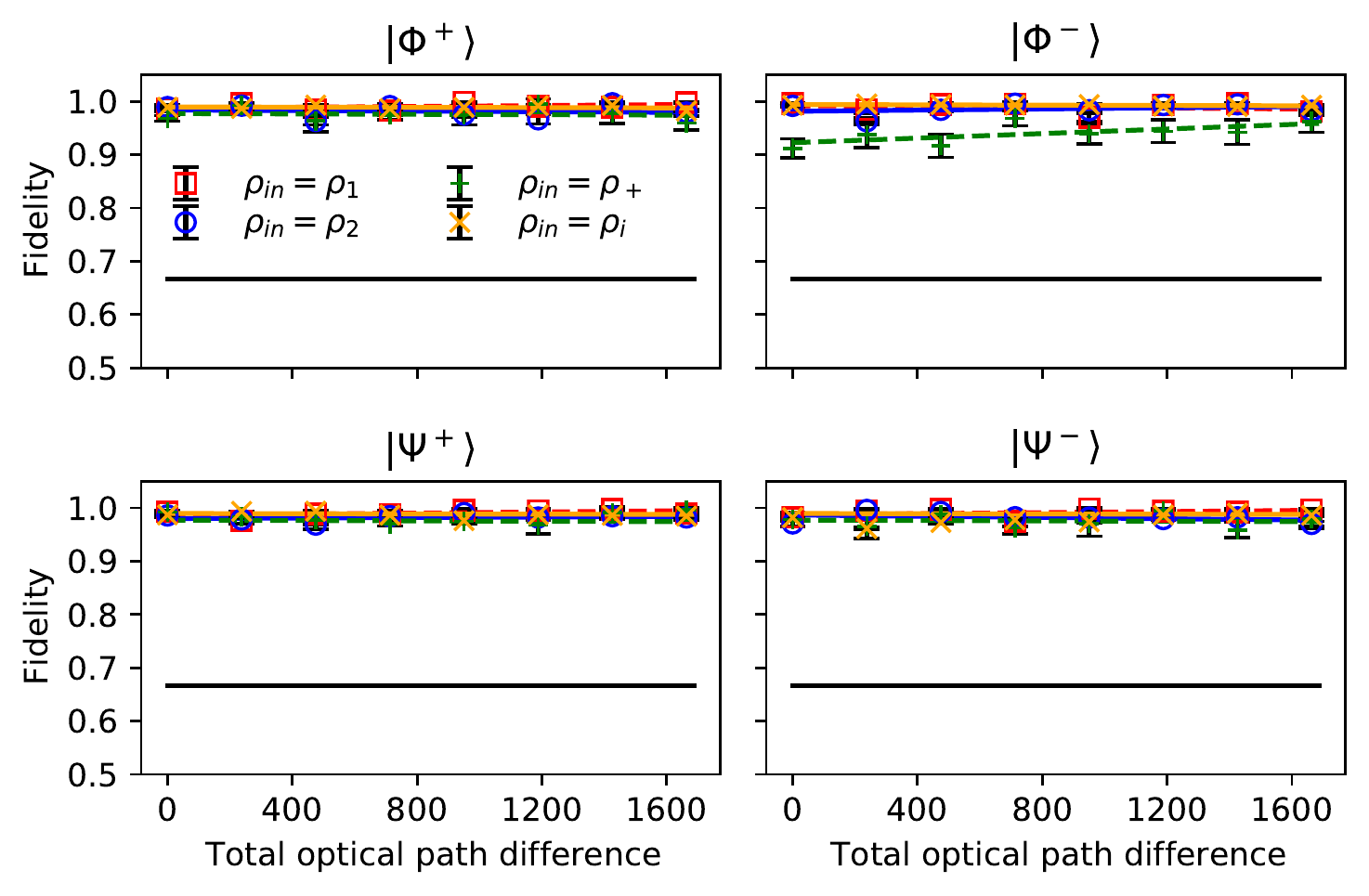}
  \caption{\label{fig:res2} The measured teleportation fidelity with error bars  as a function of increasing and equal duration of noise on both Alice's and Bob's side.
 Teleported input states $\rho_{\text{in}}$  are $\rho_{+}$ [Eq.~\eqref{eq:in1}] ("+" symbols),  $\rho_{1}$ [Eq.~\eqref{eq:in11}] (squares),
  $\rho_{2}$ [Eq.~\eqref{eq:in2}] (circles), and  $\rho_{i}$ [Eq.~\eqref{eq:ini}] ("x" symbols).
Since the earlier fits for Fig.~\ref{fig:res1} show that we are very close to the ideal case, the solid and dashed lines above are for simplicity linear fits for the experimental results. 
 The results correspond to the second set of experimental results (see the main text). The optical path difference corresponds to interaction time $t_a+t_b$ and is expressed with the unit of $808$nm. The fidelity remains essentially constant despite of having more and more noise. The black horizontal line indicates the classical limit of the fidelity with value of $2/3$.The error bars are standard deviations calculated by the Monte-Carlo method and mainly due to the counting statistics. In  most cases, the error bars are smaller than the symbols.}
  \end{center}
\end{figure}

In the first part of the first experiment,  the duration of the noise $t_a$ in Alice's side is increased in stepwise manner before her Bell measurement.
She has prepared the path qubit state to be teleported
which is given by, based on experimental state tomography as,
\begin{equation}
\label{eq:in1}
\rho_+=\left(\begin{array}{cc} 0.5507 &0.4871+i0.1007 \\0.4871-i0.1007& 0.4492\end{array}\right).
\end{equation}
This is expressed in the path qubit basis $|0\rangle$ and $|1\rangle$ and has the purity $0.999955$.
Bob, after receiving the Bell measurement outcome from Alice, implements the unitary transformation on his qubit in usual way but does not implement any noise.
The left sides of the four panels in Fig.~\ref{fig:res1} show that in this case the teleportation fidelity decreases with increasing amount of noise, as expected and going below the classical limit.
Now, in the second part of the experiment, in Alice's side there is always maximum duration of noise corresponding to optical path difference $237.6 \lambda_0$ with 
$\lambda_0=808$nm -- 
 and Bob begins to increase the duration of noise in his side in stepwise manner. 
The results are displayed on the right sides of the four panels in Fig.~\ref{fig:res1}. With the increasing amount of noise in Bob's side, the teleportation fidelity increases. Note that in all of the cases, as well as for the results below, the experimental results for fidelity are based making state tomography also for the output state.
When Bob has, in the last step, maximum duration of noise corresponding to the one on the side of Alice, the high-fidelity values correspond to the case as if there had essentially not been any noise in any part of the protocol, i.e., the ideal teleportation efficiency is recovered. For example, for Bell measurement outcome $|\Psi^+\rangle$,
the fidelity value without any noise [the first experimental point in the lower-left panel of Fig.~\ref{fig:res1}] is $0.978\pm0.003$ while fidelity value with maximum duration of noise in both sides [the last experimental point in the lower-left panel of Fig.~\ref{fig:res1})] is $0.978\pm0.006$.

In the second experiment, the duration of the local noise in both sides of Alice and Bob is increased in stepwise manner, in equal steps in both sides.
Figure~\ref{fig:res2} shows the results for four Bell measurement outcomes and using four different Alice's states to be teleported.
One of the input states is given by Eq.~\eqref{eq:in1} and the three others, based on input state tomography, are
\begin{eqnarray}
\rho_1=\left(\begin{array}{cc} 0.9794 & 0.0151-i0.0321 \\0.0151+i0.0321 & 0.0206\end{array}\right),  \label{eq:in11}\\
\rho_2=\left(\begin{array}{cc} 0.0255 & 0.01304+i0.0932  \\0.01304-i0.0932 & 0.9745\end{array}\right), \label{eq:in2} \\
\rho_i=\left(\begin{array}{cc} 0.6060 &0.02178-i0.4827  \\0.02178+i0.4827 &0.3940\end{array}\right). \label{eq:ini}
\end{eqnarray}
The purities these states are $0.962166,~0.968013$ and $0.989419$, respectively.
When there is no noise at all, the left-most points in the panels, the high-fidelity teleportation is achieved, as expected. 
However, when there is increasing duration of local noise on both sides of Alice and Bob, the fidelity does not reduce and remains essentially constant.
For example with input state $\rho_{\text{in}}= \rho_{i}$ [Eq.~\eqref{eq:ini}]and Bell-measurement result $|\Psi^+\rangle$, the fidelity without any noise is $0.987\pm0.002$ and with maximum duration of noise 
$0.985\pm0.002$.
This gives clear experimental demonstration, that even though there is hardly any entanglement left in the joint two-photon polarization state,
one can in any case achieve high-fidelity teleportation when exploiting other useful resources available when considering also other degrees of freedom and the environment of an open system.

%\textit{Discussion and conclusions.--}%
\section{Discussion and conclusions}

We have realized experimentally a scheme for high-fidelity teleportation with dephasing noise. 
Here, a pair of entangled qubits -- which  initially contain the quantum resource for the protocol -- is actually an open quantum system where each of the qubits interact with their local environments. Without the steps of the teleportation protocol, the dynamics of this bipartite open system displays non-Markovian features when the local environments of the qubits are initially correlated~\cite{nlnm,bhliu13}. Note that there also exists a quantitave connection between the amount of non-Markovianity and the teleportation fidelity~\cite{telep}.
Therefore, we have demonstrated experimentally, that it is possible to implement high-fidelity teleportation via non-Markovian open quantum system.

Our results also show that it is not necessary, that the original quantum resource resides anymore in the degrees of freedom or in the open system, which are explicitly used in the original protocol --  as long as useful resources still exist within or in the combination with the environment of the open system.
It is also worth noting here that in the described teleportation scheme, Alice's photon is destroyed in her Bell measurement while at this point Bob has not yet done anything with his qubit. Despite of this fact, Bob's subsequent open system qubit dynamics is influenced by the initially existing correlations between the two photons, even though Alices photon does not exist anymore. In a sense, in addition to teleporting a qubit state, the protocol allows to engineer in nonlocal manner -- both in time and in location --  the open system dynamics of Bob's qubit. 

In general, we have given fundamental experimental results combining concepts from open quantum systems and quantum information.
So far, there exists a number of sophisticated experiments which, e.g.,  implement reservoir engineering, quantum simulate Markovian open system dynamical maps, control Markovian to non-Markovian transition and decoherence (see e.g.~\cite{wineland,exp2,zdliu,blatt}).
However, very little is known or fundamentally tested yet when going beyond the traditional open system - environment setting, and combining the open system dynamics with sophisticated quantum information protocols or other well-known quantum physical or optical schemes including, e.g., interferometry.
We hope that our current results stimulate further work for this direction and helps to explore new areas of quantum physics where open quantum systems, and their study,
is being used outside their traditional framework.

\acknowledgments
%\textit{Acknowledgments}
This work was supported by the National Key Research and Development Program of China (No.~2017YFA0304100), 
the National Natural Science Foundation of China (Nos.~62005263, 11774335, 11821404, 11874345)
Key Research Program of Frontier Sciences, CAS (No.~QYZDY-SSW-SLH003), the Fundamental Research Funds for the Central Universities (No.~WK2470000026), Science Foundation of the CAS (No.~ZDRW-XH-2019-1), Anhui Initiative in Quantum Information Technologies (AHY020100), and Science and Technological Fund of Anhui Province for Outstanding
Youth (No.~2008085J02). 
Z. D. L acknowledges financial support from China Postdoctoral Science Foundation (Grant No. 2020M671862).
S. H. R. acknowledges financial support from Finnish Cultural Foundation and Turku University Foundation.
J. P. acknowledges financial support from Magnus Ehrnrooth Foundation.

\end{document}